\documentclass[a4paper,fleqn,usenatbib]{mnras}
\usepackage{newtxtext,newtxmath,hyperref}
\usepackage[T1]{fontenc}
\usepackage{ae,aecompl}
\usepackage[dvipdfmx]{graphicx}
\usepackage{amsmath}
\usepackage{amssymb}
\usepackage{multirow,cases,colortbl}
\usepackage{url}
\usepackage{bm}
\usepackage{lscape}
\usepackage{ulem,color}

\newcommand{\MP}{\ensuremath{m_{\rm{p}}}}
\newcommand{\ME}{\ensuremath{m_{\rm{e}}}}

\newcommand{\gr}{\ensuremath{\gamma}-ray }
\newcommand{\grs}{\ensuremath{\gamma}-rays }

\title[Relativistic origin of long-lasting emission of SGRBs]{Linking extended and plateau emissions of short gamma-ray bursts}

\author[Matsumoto et al.]{Tatsuya Matsumoto,$^{1,2,3,4}$
Shigeo S. Kimura,$^{4,5,6,7,8,9}$
Kohta Murase,$^{5,6,7,10}$
\newauthor and Peter M\'{e}sz\'{a}ros$^{5,6,7}$
\\
$^{1}$Racah Institute of Physics, Hebrew University, Jerusalem, 91904, Israel\\
$^{2}$Research Center for the Early Universe, Graduate School of Science, University of Tokyo, Tokyo 113-0033, Japan\\
$^{3}$Department of Physics, Graduate School of Science, University of Tokyo, Tokyo 113-0033, Japan\\
$^{4}$JSPS Research Fellow\\
$^{5}${Department of Physics, The Pennsylvania State University, University Park, Pennsylvania 16802, USA}\\
$^{6}${Department of Astronomy \& Astrophysics, The Pennsylvania State University, University Park, Pennsylvania 16802, USA}\\
$^{7}${Center for Particle and Gravitational Astrophysics, The Pennsylvania State University, University Park, Pennsylvania 16802, USA}\\
$^{8}${Frontier Research Institute for Interdisciplinary Sciences, Tohoku University, Sendai 980-8578, Japan}\\
$^{9}${Astronomical Institute, Tohoku University, Sendai 980-8578, Japan}\\
$^{10}${Yukawa Institute for Theoretical Physics, Kyoto, Kyoto 606-8502, Japan}
}

\pubyear{2019}

\begin{document}
\label{firstpage}
\pagerange{\pageref{firstpage}--\pageref{lastpage}}
\maketitle

\begin{abstract}
Some short gamma-ray bursts (SGRBs) show a longer lasting emission phase, called extended emission (EE) lasting $\sim 10^{2-3}\,\rm s$, as well as a plateau emission (PE) lasting $\sim10^{4-5}\,\rm s$.
While a long-lasting activity of the central engines is a promising explanation for powering both emissions, their physical origin and their emission mechanisms are still uncertain.
In this work, we study the properties of the EEs and their connection with the PEs.
First, we constrain the minimal Lorentz factor $\Gamma$ of the outflows powering EEs, using compactness arguments and find that the outflows should be relativistic, $\Gamma\gtrsim10$.
We propose a consistent scenario for  the PEs, where the outflow eventually catches up with the jet responsible for the prompt emission, injecting energy into the forward shock formed by the prior jet, which naturally results in a PE.
We also derive the radiation efficiency of EEs and the Lorentz factor of the outflow within our scenario for  10 well-observed SGRBs accompanied by both EE and PE.
The efficiency has an average value of $\sim3\,\%$ but shows a broad distribution ranging from $\sim0.01$ to $\sim100\%$.
The Lorentz factor is $\sim20-30$, consistent with the compactness arguments.
These results suggest that EEs are produced by a slower outflow via more inefficient emission than the faster outflow which causes the prompt emission with a high radiation efficiency.
\end{abstract}


\section{Introduction}
Short gamma-ray bursts (SGRBs) are a subclass of GRBs, whose duration ($T_{90}$) is shorter than 2~s \citep[see][for reviews]{Nakar2007,Berger2014}.
A progenitor of SGRBs has been considered to be the coalescence of a compact binary merger including at least a neutron star \citep[NS,][]{Goodman1986,Paczynski1986,Eichler+1989}. 
The recent detection of gravitational waves (GWs) and electromagnetic counterparts from a binary NS merger, GW170817 \citep{Abbott+2017c,Abbott+2017_gw170817_multi}, supports this idea.
The VLBI observations revealed a super-luminal motion of a radio point source, which is well explained by an emission from a relativistic jet \citep{Mooley+2018b,Ghirlanda+2019}.
It should be noted that the \gr counterpart, GRB 170817A \citep{Abbott+2017e,Goldstein+2017,Savchenko+2017}, is unlikely to be emitted by a core of the jet \citep[see, e.g.,][for such an off-axis scenario]{Ioka&Nakamura2018}, which is usually observed for regular SGRBs, and this GRB is not decisive evidence that the binary NS merger produced a SGRB \citep{Kasliwal+2017,Matsumoto+2019,Matsumoto+2019b}.

One of the biggest puzzles on SGRBs is what powers long-lasting emissions in early-time X-ray afterglows.
Such extended components are first recognized in the observations of GRBs 050709 and 050724 \citep{Barthelmy+2005c,Villasenor+2005}.
This component, whose typical duration and luminosity are $\sim10^{2-3}\,\rm s$ and $\sim10^{47-49}\,\rm erg\,s^{-1}$, is called an extended emission (EE).
Systematic analyses of afterglow emission show that some fraction or even most of SGRBs are accompanied with EEs \citep{Norris&Bonnell2006,Sakamoto+2011,Gompertz+2013,Kagawa+2015,Kisaka+2017}.
In particular, EEs show rapid time variability or decay, which suggests that they are powered by prolonged activities of a central engine \citep[e.g.,][]{Ioka+2005}.
Thus, EEs may be powered by dissipation of an outflow launched by a merger remnant such as a black hole \citep[BH,][]{Barkov&Pozanenko2011,Nakamura+2014,Kisaka&Ioka2015} or a highly magnetized neutron star (NS) \citep[magnetar,][]{Metzger+2008,Bucciantini+2012,Rowlinson+2013,Gompertz+2013,Gompertz+2014,Gompertz+2015,Lu+2015,Rezzolla&Kumar2015,Gibson+2017}, while the origin and the emission mechanism of EEs are still uncertain.
 
Recently, another plateau component has been identified for some GRBs with EEs \citep{Gompertz+2013,Kisaka+2017}. 
This additional plateau typically continues for a longer time of $\sim10^{4-5}\,\rm s$ with a lower luminosity of $\sim10^{44-46}\,\rm erg\,s^{-1}$ than the EEs \citep{Kisaka&Ioka2015,Kisaka+2017}.
We call this component as a plateau emission (PE), but it should be noted that some authors call it as a late-time plateau, an X-ray plateau \citep{Gompertz+2014}, or an external plateau \citep{Lu+2015}.\footnote{\cite{Lu+2015} classify plateaus of SGRB's afterglows into internal and external plateaus based on their temporal decay: the internal plateau declines faster than $t^{-2}$ and the external one does slower. On the other hand, we empirically identify EEs and PEs mainly according to their typical timescales \citep[e.g.,][]{Kisaka+2017}. Internal plateaus are identical with EEs as discussed by \citealt{Lu+2015} and external plateaus are also likely to correspond to PEs because their observable quantities are similar to those of PEs \citep{Lu+2015}.}
In addition to EEs, as in plateau and shallow-decay emissions of long GRBs \citep[][]{Ghisellini+2009,Murase+2011}, PEs have often been attributed to the central engine activities (\citealt{Gompertz+2013,Gompertz+2014,Gompertz+2015,Kisaka&Ioka2015}, but see e.g., \citealt{Oganesyan+2019,Beniamini+2019c} for different ideas).

\begin{figure}
\begin{center}
\includegraphics[width=110mm, angle=90]{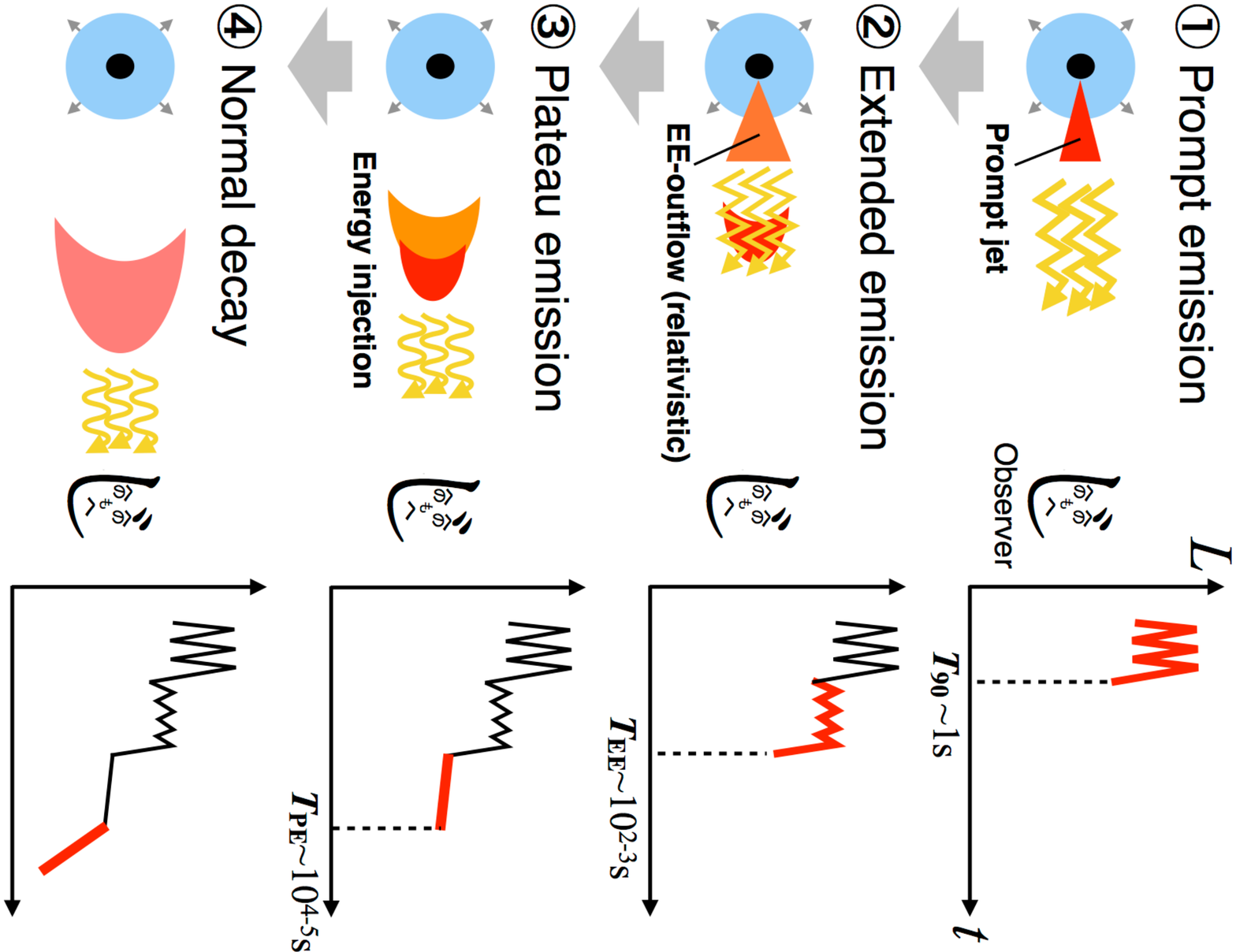}
\caption{Schematic picture of our scenario for EE and PE. (1) A relativistic prompt jet produces prompt emission. 
(2) The central engine continues to be active and launch a slower outflow, produces an EE. 
The EE-outflow can be either connected with or disconnected from the prompt jet, but its Lorentz factor is smaller than that of the prompt jet. 
(3) The EE-outflow catches up with the decelerating prompt jet and injects energy into an external shock, which results in a PE from a forward shock. 
(4) Finally, the merged outflow evolves adiabatically and produces a standard afterglow (normal decay).}
\label{fig picture}
\end{center}
\end{figure}

In this work, we constrain the minimal Lorentz factor of outflows which produce EEs, and propose that they are causally connected to PEs.
First, we derive the minimal Lorentz factor of EEs by using compactness arguments, which are usually applied to prompt emissions \citep[e.g.,][]{Lithwick&Sari2001}.
We find that outflows producing EEs should be relativistic with a Lorentz factor of $\gtrsim10$.
Then, we consider a possible connection between EEs and PEs and propose their consistent picture. 
In Fig. \ref{fig picture}, we show a schematic picture of our scenario, where a relativistic outflow which produces an EE catches up with a decelerating prompt jet, and injects energy into an external shock.
Due to the energy injection, the emission from the shock produces a shallow decay \citep[e.g.,][]{Rees&Meszaros1998,Kumar&Piran2000,Sari&Meszaros2000,Zhang+2006,Ioka+2006}, which we observe as a PE.
In this scenario, PEs are a natural outcome of the interaction between the outflows producing prompt emissions and EEs.
Finally, we evaluate the emission efficiency and Lorentz factor of the outflow producing EEs within our model.

We organize this paper as follow.
In \S \ref{compactness}, we constrain the Lorentz factor of the outflow producing EEs by compactness considerations.
We turn to focus on the behavior of X-ray afterglow following the EEs in \S \ref{scenario}.
We propose a scenario of the X-ray afterglow including PEs.
Within this model, we calculate the emission efficiency of EEs and Lorentz factor of the outflow producing EEs.
In \S \ref{summary}, we summarize this work and discuss implications of our results for emission mechanisms of EEs and central engines.

\section{Minimal Lorentz factor of outflows producing extended emission}\label{compactness}
We derive the minimal Lorentz factor of EEs based on compactness considerations.
Compactness arguments are a powerful tool to constrain the minimal Lorentz factor of prompt emissions in GRBs \citep{Krolik&Pier1991,Lithwick&Sari2001,Matsumoto+2019b}.
In these arguments, it is usually required that a detected most energetic photon escapes from a \gr emitting site without producing an electron-positron pair by colliding with other photons (this condition is named as limit A by \citealt{Lithwick&Sari2001}).
While this condition gives a stringent limit on the Lorentz factor for luminous and variable prompt emissions, it is not so constraining for dimmer emissions such as EEs.
Thus, we consider the other two conditions and call them limits B and C according to \cite{Lithwick&Sari2001}.

In limit B, the Compton scattering by produced pairs is the main opacity source. 
Using an observed spectrum, we evaluate the number of pairs which is assumed to be equal with the number of photons with larger energy than the threshold for self-annihilation in the observer frame: 
\begin{align}
\varepsilon_{\rm th}=2\Gamma\ME c^2,
\end{align}
where $\Gamma$, $\ME$, and $c$ are the Lorentz factor of the emitting site (the outflow producing EEs), electron mass, and speed of light, respectively.
As shown in Table \ref{table list}, most of EEs are consistent with a single power-law (PL) spectrum: $dN/d\varepsilon \propto \varepsilon^{\alpha_p}$, where $\alpha_p$ is the photon index.
To be conservative, we introduce an exponential cutoff in the spectrum as the maximal photon energy of a detector's energy band, $\varepsilon_{\rm max}$: 
\begin{align}
\frac{dN}{d\varepsilon} \propto \varepsilon^{\alpha_p}\exp\biggl(-\frac{\varepsilon}{\varepsilon_{\rm max}}\biggl).
\end{align}
The number fraction of photons $f$, which can produce pairs by the self-annihilation of \grs, is evaluated by
\begin{align}
f=\frac{\int_{\varepsilon_{\rm th}}^{\infty}d\varepsilon\,\varepsilon^{\alpha_p}\exp\big(-\frac{\varepsilon}{\varepsilon_{\rm max}}\big)}{\int_{\varepsilon_{\rm min}}^{\infty}d\varepsilon\,\varepsilon^{\alpha_p}\exp\big(-\frac{\varepsilon}{\varepsilon_{\rm max}}\big)}=\frac{\int_{\varepsilon_{\rm th}/\varepsilon_{\rm max}}^\infty dx\,x^{\alpha_p}e^{-x}}{\int_{\varepsilon_{\rm min}/\varepsilon_{\rm max}}^\infty dx\,x^{\alpha_p}e^{-x}},
   \label{eq fraction b}
\end{align}
where $\varepsilon_{\rm min}$ is the minimal energy of the detector and the denominator is the normalization of the spectrum.\footnote{In \cite{Matsumoto+2019,Matsumoto+2019b}, the normalization of spectra is ignored. This is justified for detectors whose ratio of maximal to minimal energy is larger than $\varepsilon_{\rm min}/\varepsilon_{\rm max}\gtrsim0.1$ as BAT and GBM. However, detectors with a smaller ratio like BATSE, the normalization should be taken into account to avoid overestimating the number fraction.}
For BATSE, BAT, and GBM, the maximal and minimal energy are given by $(\varepsilon_{\rm min},\,\varepsilon_{\rm max})$ = (30, 2000), (15, 150), and (50, 300), respectively, in unit of keV.
For a burst with a cutoff power-law (CPL) spectrum with $\alpha_p>-2$, the maximal energy is replaced with $\varepsilon_{\rm pk}/(\alpha_p+2)$, where $\varepsilon_{\rm pk}$ is the spectral peak energy.

In limit C, the Compton scattering by electrons associated with baryons in the outflow is the opacity source.
It should be noted that limit C is the same as the condition for baryonic photospheres and relevant only for the baryon-dominated outflow (which implies that it does not work for Poynting-flux-dominated outflows).
The number of electrons is estimated by energetics, that is the observed total photon energy should be smaller than the baryon kinetic energy.
As we show below, we can use the same functional form for the optical depth as limit B, by using the ``number fraction'' given by \citep{Matsumoto+2019b}
\begin{align}
f=\frac{\varepsilon_{\rm min}}{2\Gamma^2\MP c^2},
   \label{eq fraction c}
\end{align}
where $\MP$ is the proton mass and we use $\varepsilon_{\rm min}$ to evaluate the total photon energy.\footnote{This prescription underestimates the minimal Lorentz factor for bursts with a hard single power-law spectrum $\alpha_p>-2$, but the resulting error is less than a factor of 2.}

The Thomson optical depth is given by
\begin{align}
\tau_{\rm T}&\simeq\frac{\sigma_{\rm T}Lf}{16\pi c^2\varepsilon_{\rm min}\delta t\Gamma^4}\simeq\frac{\sigma_{\rm T}Sd_{\rm L}^2f}{4c^2\varepsilon_{\rm min}\delta t T \Gamma^4},
   \label{eq tau}
\end{align}
where $\sigma_{\rm T}$, $L(=4\pi d_{\rm L}^2 S/T)$, $d_{\rm L}$, $S$, $T$, and $\delta t$ are the Thomson cross section, luminosity, luminosity distance, observed fluence, total duration, and variable timescale of EEs, respectively.
The number fraction of photons in each limit is given by Eqs. \eqref{eq fraction b} and \eqref{eq fraction c}, and summarized as
\begin{align}
f&=\begin{cases}
\frac{{\bf \Gamma}(\alpha_p+1,\varepsilon_{\rm th}/\varepsilon_{\rm max}(1+z))}{{\bf \Gamma}(\alpha_p+1,\varepsilon_{\rm min}/\varepsilon_{\rm max})}&\text{; Limit B,}\\
\frac{\varepsilon_{\rm min}(1+z)}{2\Gamma^2\MP c^2}&\text{; Limit C,}
\end{cases}
\end{align}
where ${\bf \Gamma}(\alpha_p+1,x)$ is the incomplete Gamma function, and we restore the redshift dependence.
In Eq. \eqref{eq tau}, we also use $\varepsilon_{\rm min}$ to estimate the total photon number.
The variability timescale is assumed to be $\delta t=1\,\rm s$ for all calculations in this paper \citep[see, e.g.,][for the light curves of EEs]{Norris&Bonnell2006}. 

We derive the minimal Lorentz factor by setting $\tau_{\rm T}=1$.
For limit B, we solve Eq. \eqref{eq tau} numerically.
For limit C, we can obtain the minimal Lorentz factor of EEs analytically,
\begin{align}
\Gamma_{\rm min}&=\biggl(\frac{\sigma_{\rm T}Sd_{\rm L}^2(1+z)}{8\MP c^4\delta tT}\biggl)^{1/6}
   \label{gamma_min c}\\
&\simeq4.4\,S_{-7}^{1/6}T_{2}^{-1/6}\delta t_0^{-1/6}\biggl(\frac{d_{\rm L}}{2.9\,\rm Gpc}\biggl)^{1/3}\biggl(\frac{1+z}{1.5}\biggl)^{1/6},
\end{align}
where we use the convention of $Q_x=Q/10^x$ (cgs).
For EEs without measured redshifts, we adopt $z=0.5$ ($d_{\rm L}=2.9\,\rm Gpc$) which is the median redshift for SGRBs with measured redshifts \cite[e.g.,][]{Berger2014}.

In Table \ref{table list}, we show the sample of EEs which we analyze using the compactness arguments.
The events are taken from \cite{Bostanci+2013,Kaneko+2015}, which contain the spectral information.
They are divided into three groups detected by BATSE, BAT, and GBM, respectively.
For BATSE GRBs, we include GRB 950531 whose $T_{90}=3.52\,\rm s$ into our sample.

We show the minimal Lorentz factors of EEs derived for limits B and C in the last column of Table \ref{table list}.
These Lorentz factors are obtained by using the central values of the observables.
The more constraining limit is shown with boldface.
All EEs should involve a relativistic motion with $\Gamma\gtrsim10$.
The minimal factor for limit C does not change so much for event by event because the Lorentz factor weakly depends on the observables (see Eq. \ref{gamma_min c}).
On the other hand, limit B gives larger minimal Lorentz factors for EEs detected by BATSE than those detected by the other two detectors.
It should be noted that the minimal value for limit B depends on the spectrum, and in particular we cut off the spectra exponentially above the maximal energy of the detectors.
Larger $\varepsilon_{\rm max}$ gives larger photon fraction (see Eq. \ref{eq fraction b}), which results in larger $\Gamma_{\rm min}$.
Thus, EEs detected by BATSE with $\varepsilon_{\rm max}=2000\,\rm keV$ can have larger $\Gamma_{\rm min}$ than those detected by BAT and GBM.

\begin{figure}
\begin{center}
\includegraphics[width=85mm, angle=0]{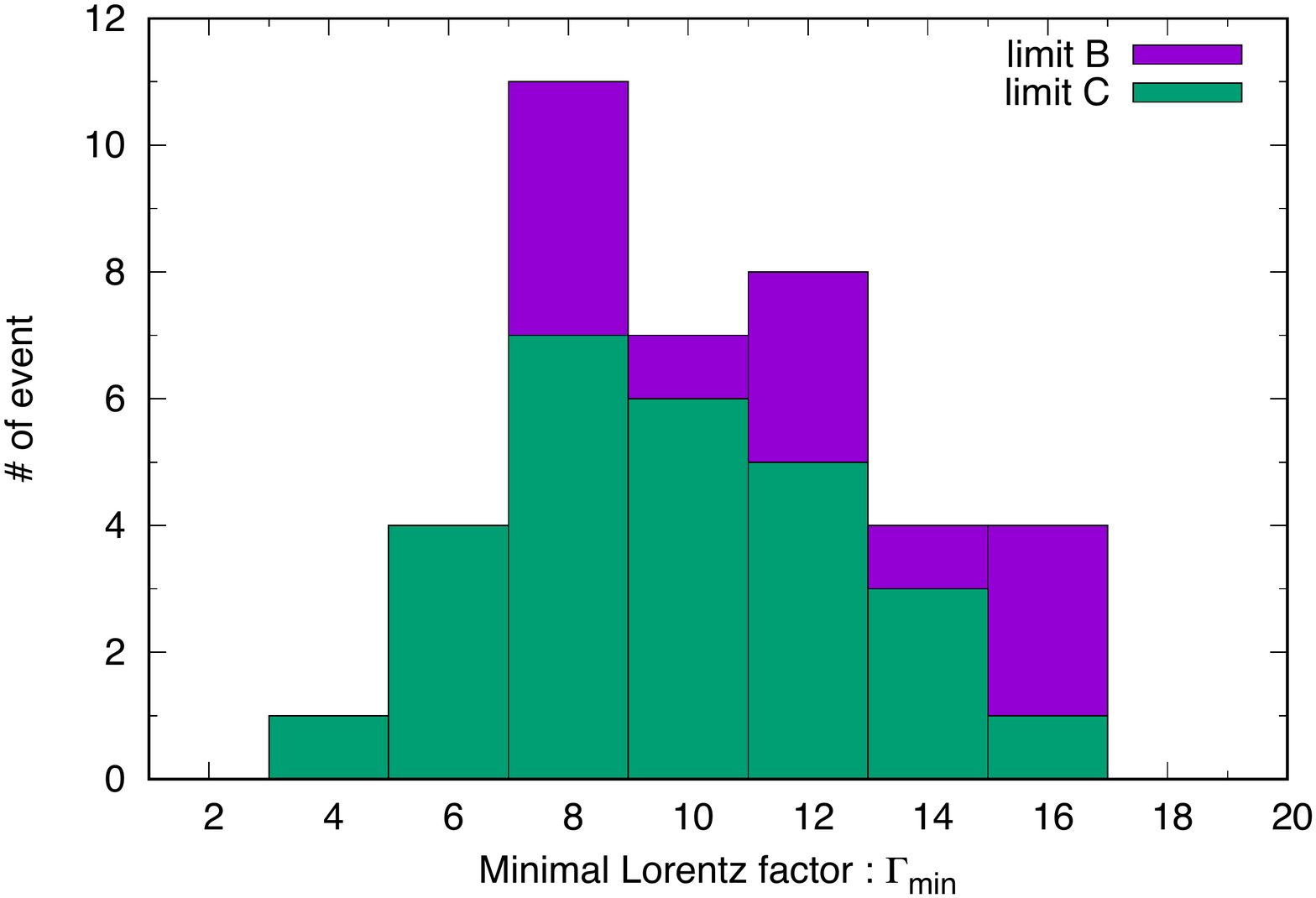}
\caption{Distribution of the minimal Lorentz factor of EEs, $\Gamma_{\rm min}$, obtained by compactness arguments. The mean and standard deviation are $10.3$ and $3.2$, respectively.}
\label{fig gamma}
\end{center}
\end{figure}

Fig. \ref{fig gamma} depicts the distribution of $\Gamma_{\rm min}$.
The mean and standard deviation of the distribution are $10.3$ and $3.2$, respectively.
Again, this is direct evidence that outflows producing EEs should be relativistic with at least $\Gamma\sim10$.
This fact not only constrains models for EEs but also motivates us to consider the fate of the relativistic outflows.

\section{Energy injection from extended-emission outflow: natural origin of plateau emission}\label{scenario}
In the previous section, we find that an EE should be produced by a relativistic outflow with $\Gamma\gtrsim10$.
Hereafter, we call outflows producing EEs as EE-outflows.

\subsection{Plateau emission powered by EE-outflows}
We consider the fate of an EE-outflow after producing an EE.
It is reasonable to consider that the Lorentz factor of the EE-outflow is smaller than that of the prompt jet mainly because EEs are much dimmer and softer than prompt emissions.
We do not specify whether the EE-outflow is connected or separated from the prompt jet, but it has a different (smaller) Lorentz factor and emission timescale from the prompt jet.
Then, such an EE-outflow finally catches up with the prompt jet and injects energy into the external shock formed by the prompt jet, which results in a plateau as originally proposed for long GRBs with shallow-decaying X-ray afterglows \citep{Rees&Meszaros1998,Kumar&Piran2000,Sari&Meszaros2000,Nousek+2006,Zhang+2006,Ioka+2006}.

Interestingly, several SGRBs with EEs actually show PEs in X-ray afterglows with a duration of $T_{\rm PE}\sim10^{4-5}\,\rm s$ \citep{Gompertz+2013,Gompertz+2014}.
Although these PEs are less luminous and difficult to obtain detailed light curves, perhaps, PEs are common in SGRBs accompanied by EEs \citep{Kisaka+2017}.
In addition to EEs, PEs are sometimes also interpreted as being powered by a prolonged central-engine activity.
In these models, the PEs are produced by the spin down activities of a magnetar \citep{Gompertz+2013,Gompertz+2014,Gibson+2017} or a jet launched from a BH \citep{Kisaka&Ioka2015} after producing an EE.

Contrary to such internal dissipation scenarios, we here propose a scenario for the emission mechanism of PEs, in which an EE-outflow with $\Gamma\gtrsim10$ collides with a decelerating prompt jet and injects energy to power a PE via an external shock.
Fig. \ref{fig picture} gives a schematic picture of our scenario.
In this scenario, (1) A prompt jet with a Lorentz factor $\Gamma_0>\Gamma$ is launched from a central engine and produces a prompt emission.
(2) The engine remains active for a long time and launches an EE-outflow with $\Gamma\gtrsim10$ which produces an EE by some internal dissipation process (which we do not specify) for $T_{\rm EE}\sim10^{2-3}\,\rm s$.\footnote{The EE-outflow does not have to be disconnected from the prompt jet. The central engine can continue to launch a jet for a long time but reduces its luminosity and Lorentz factor.}
(3) While the prompt jet sweeps up interstellar medium (ISM) and decelerates, the EE-outflow coasts and catches up with the prompt jet.
By the subsequent collision, the EE-outflow injects energy into the prompt-jet shell and changes the dynamics of the shell from an adiabatic evolution and produces a shallow decay in the afterglow \citep{Rees&Meszaros1998,Kumar&Piran2000,Sari&Meszaros2000}, which is observed as a PE.
It should be stressed that in the present model, the PE is produced not by long-lasting engine activities for $T_{\rm PE}$, but by the (external) forward shock with the energy injection, which is a natural outcome if the EE is produced by a relativistic outflow.
(4) After the whole EE-outflow collides with the shell at $T_{\rm PE}\sim10^{4-5}\,\rm s$, the merged outflow expands adiabatically to produces a standard afterglow (normal decay phase).

It should be noted that two types of energy injection mechanisms have originally been discussed for long GRBs \citep[e.g.,][]{Zhang+2006}.
One is the long-lived engine model \citep{Dai&Lu1998,Zhang&Meszaros2001}, where an engine continues to launch an outflow for longer time than the duration of prompt emissions and the outflow collides with a prompt jet and injects energy.
The other is the delayed outflow model \citep{Rees&Meszaros1998,Sari&Meszaros2000}, where the engine shuts down after launching the prompt jet, but the prompt outflow has a range of Lorentz factors and the slower outflow eventually catches up with the faster one.
In either case a forward shock is refreshed.
In the context of our PE scenario, the injection process may be well described by the delayed outflow model because the launching time of the EE-outflow should be comparable to the duration of EEs.
Thus, the lifetime of the central engine is reasonably assumed to be short compared with the PE timescale.

The interpretation of PEs as emission from a refreshed shock by an EE-outflow has several advantages over the late dissipation scenario, where the emission is attributed to internal emission associated with the long-lasting central engine.
First, a PE is naturally explained as a result of the interaction between an EE-outflow and a prompt jet.
In the late dissipation scenario, both EE and PE are explained by complicated central-engine activities with different timescales.
For example, in a magnetar model \citep{Gompertz+2014,Gibson+2017}, it is considered that EE and PE are powered by the propeller effect and spin-down activities of magnetars, respectively, while it is unclear whether a relativistic EE-outflow can be launched by the propeller effect or not.
Among BH models, \cite{Kisaka&Ioka2015} proposed that both emissions are produced by a BH jet launched by Blandford-Znajek process \citep{Blandford&Znajek1977}, where the fallback accretion and magnetic reconnection is supposed to make two plateaus in the jet power.
However, such a long-time evolution of fallback matter or magnetic field is still uncertain \cite[e.g.,][for a possible effect modifying the fallback accretion]{Desai+2019}.

Second, the refreshed shock model for PEs can explain the observed smooth transition from a PE to a normal decay better than the long-lived engine models.
As we see in a sample of PEs in Fig. \ref{fig lc}, most of the PEs smoothly connect with normal decays while EEs decay rapidly.
Originally, such a rapid shutdown, which external shocks cannot produce, required to consider central engines as an origin of EEs \citep{Ioka+2005}.
Thus, basically there is no positive motivation to relate PEs directly with central-engine activities.
Furthermore, spectra of PEs are similar to those of normal decay.
Recently, \cite{Zhao+2019} carried out temporal and spectral analyses of PEs and normal decays for a large sample of GRBs including SGRBs.
They confirmed that as \cite{Liang+2007} found, the spectral indexes do not change between both phases, and consistent with the simple energy injection scenario.

In addition, the well-observed GRB 060614 shows an achromatic break at the end of the PE, which is predicted by the delayed energy injection scenario \citep{Mangano+2007}.
On the other hand, it is uncertain whether engine models can reproduce such a spectrum of PE, where the PE is produced by an internal dissipation of an outflow.
These observational facts, a smooth connection and the same spectrum with normal decays, support that PEs and normal decays are produced in the same emitting region, that is an external forward shock.

Finally, we comment on several works where a similar possibility for PEs was discussed.
\cite{Gompertz+2015} considered that a PE is powered from a forward shock with an energy injection from the spin-down activity of a magnetar, although they did not relate the injection with EEs.
Similarly, \cite{Lu+2015} also discussed the origin of external plateaus to be an energy injection.
For GRBs 060614 and 160821B, which are observed in detail, \cite{Xu+2009,Lamb+2019} constructed multi-wavelength models, respectively.
Both of them identified EEs and PEs in afterglows and modeled PEs in light of the energy injection scenario (but see, \citealt{Troja+2019b} for a different interpretation of GRB 160821B).
In particular, they considered that the energy injection results from the interaction between the prompt jet and EE-outflows as we consider in this work.
Their detailed modeling supports that our picture holds for these specific events and it is natural for us to extend this model to all SGRBs with EEs and PEs.

\subsection{Radiation efficiency and Lorentz factor of EE-outflows}
In our scenario, EEs are causally connected to PEs, and we can constrain the properties of EEs by observing PEs.
We here focus on the radiation efficiency and Lorentz factor of EE-outflows.
The radiation efficiency is introduced as the ratio of radiation energy to kinetic energy as
\begin{align}
\eta_{\rm EE}\equiv\frac{E_{\rm EE,iso}}{E_{\rm EE,iso}+E_{\rm k,iso}},
\label{eq eta}
\end{align}
where $E_{\rm EE,iso}$ is the isotropic-equivalent radiation energy of the EE.
In our model, the isotropic kinetic energy of the EE, $E_{\rm k,iso}$ is given by analyzing the normal decay phase because the energy injected by the EE-outflow into the shell dominates the total kinetic energy.

The Lorentz factor of the slowest part of the EE-outflow is estimated by the timescale of the PE because the PE ends when the whole EE-outflow collides with a shell formed by a prompt jet.
In our scenario, after producing the EE, the EE-outflow has a range of Lorentz factor as the delayed outflow model for the energy injection.
A part of the outflow with a Lorentz factor $\Gamma$ collides with the prompt-jet shell when the shell's Lorentz factor becomes comparable with $\Gamma$.\footnote{More accurately, this collision occurs when the Lorentz factor of the EE-outflow is a few times larger than the shell's Lorentz factor in the delayed outflow scenario \citep{Kumar&Piran2000}.}
We denote the Lorentz factor of (shocked) EE-outflow after the collision as $\Gamma_{\rm EE}(<\Gamma)$.
Thus the Lorentz factor of the slowest part of the EE-outflow is evaluated by the Lorentz factor of the merged outflow at the beginning of the normal decay phase, which is evaluated by \citep{Sari+1998}
\begin{align}
\Gamma_{\rm EE}\simeq24\,\biggl(\frac{T_{\rm PE}/(1+z)}{10^{4}{\,\rm s}}\biggl)^{-3/8}\biggl(\frac{E_{\rm k,iso}}{10^{52}\,\rm erg}\biggl)^{1/8}\biggl(\frac{n}{10^{-2}\,\rm cm^{-3}}\biggl)^{-1/8},
\label{eq gamma ee}
\end{align}
where $n$ is the number density of ISM.
It should be noted that practically another part of the (unshocked) EE-outflow can have a larger Lorentz factor than $\Gamma_{\rm EE}$.
However, we cannot estimate its Lorentz factor from the PE timescale because such part has already caught up with the shell.

\begin{figure*}
\begin{center}
\includegraphics[width=165mm, angle=0]{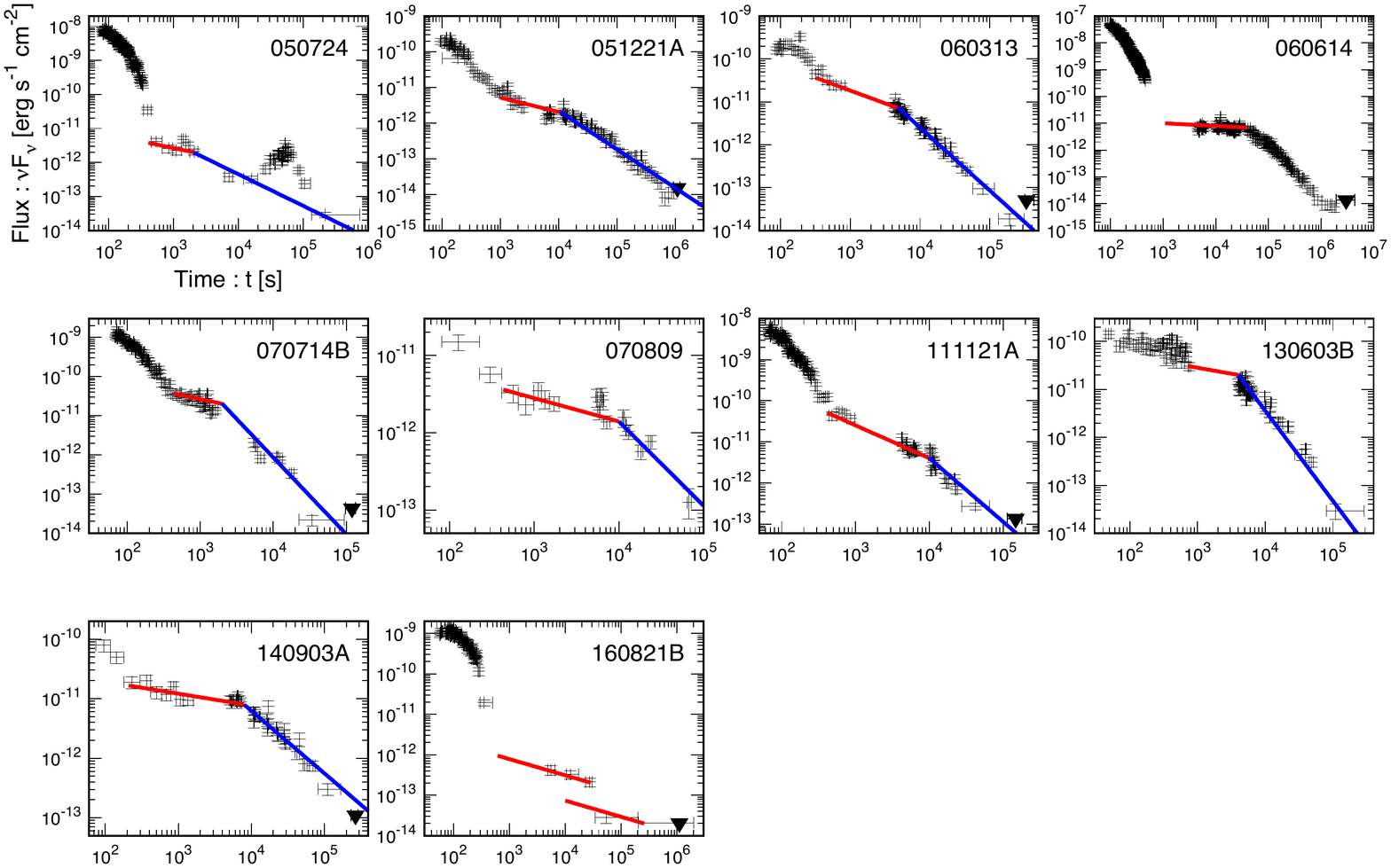}
\caption{X-ray light curves of SGRBs in our sample (listed in Table \ref{table list2}) taken from the XRT catalog ($0.3-10\,\rm keV$). Red and blue lines show PEs and normal decays, respectively. The blue lines are shown when the temporal indexes are given by \citealt{Fong+2015}.}
\label{fig lc}
\end{center}
\end{figure*}

In Table \ref{table list2}, we list SGRBs with EEs and PEs, which are taken from \cite{Kisaka+2017,Kagawa+2019}.
Although most of SGRBs are consistent with the picture that they have both EE and PE \citep{Kisaka+2017}, it is difficult to obtain detailed light curves of PEs due to the limited sensitivity of detectors. 
Therefore, to be conservative, we consider only 10 SGRBs showing evident PEs from the references.
Fig. \ref{fig lc} depicts the X-ray light curves of the sample seen by XRT.\footnote{\url{https://www.swift.ac.uk/index.php}}
The PEs and normal decays are shown with red and blue lines, respectively.
We overlay the red lines and {\it ad hoc} connect them to the blue ones when their temporal indexes are given by \cite{Fong+2015}.

\begin{table*}
\begin{center}
\caption{List of SGRBs showing EE and evident PE (taken from \citealt{Kisaka+2017,Kagawa+2019}). 
Spectral indices at BAT- ($15-150\,\rm keV$) and XRT-band ($0.2-10\,\rm keV$) are taken from \citealt{Kagawa+2019} (see their figure 1). $E_{\rm EE,iso}$ is also taken from \citealt{Kagawa+2019} or calculated by ourselves taking the energy correction to $2-150\,\rm keV$ range into account (see the text). $E_{\rm k,iso}$ is taken from references, where the energy is obtained by fitting normal decays by the standard afterglow model because the energy is dominated by injected energy by the EE-outflow within our model.}
\label{table list2}
\begin{tabular}{lrlrrrrrrrr}
\hline
Event&(Ref)&Redshift&\multicolumn{2}{r}{Spectral index}&$E_{\rm EE,iso}$ (2-150 keV)&$E_{\rm k,iso}$&$n$&$T_{\rm PE}$&$\eta_{\rm EE}$&$\Gamma_{\rm EE}$\\
&&$z$&BAT&XRT&[erg]&[erg]&[$\rm cm^{-3}$]&[s]&&\\
\hline\hline
050724&(1)&0.257&$-2.00^{+0.26}_{-0.27}$&$-1.34^{+0.13}_{-0.23}$&$2.0\times10^{51}$&$1.5\times10^{51}$&$1\times10^{-1}$&$\sim2\times10^{3}$&$5.7\times10^{-1}$&28\\
051221A&(2)&0.547&-&$-1.85^{+0.13}_{-0.21}$&$1.9\times10^{50}$&$\sim1\times10^{51}$&$\sim1\times10^{-3}$&$1.1\times10^4$&$1.3\times10^{-1}$&28\\
060313&(3)&(0.5)&-&$-1.07^{+0.13}_{-0.23}$&$\sim3\times10^{50}$&$4.5\times10^{51}$&$3.3\times10^{-3}$&$\sim5\times10^3$&$6.3\times10^{-2}$&38\\
060614&(4,5)&0.125&$-2.35^{+0.04}_{-0.04}$&$-0.82^{+0.05}_{-0.05}$&$4.3\times10^{51}$&$2.3\times10^{52}$&$4\times10^{-2}$&$3.7\times10^{4}$&$1.6\times10^{-1}$&14\\
070714B&(3)&0.923&$-2.06^{+0.27}_{-0.29}$&$-0.98^{+0.11}_{-0.19}$&$7.8\times10^{51}$&$1.0\times10^{51}$&$5.6\times10^{-2}$&$\sim2\times10^{3}$&$8.9\times10^{-1}$&34\\
070809&(3)&0.219&-&-&$2.9\times10^{48}$&$\sim7\times10^{51}$&$\sim5\times10^{-5}$&$\sim1\times10^{4}$&$3.9\times10^{-4}$&48\\
111121A&(3)&($0.5$)&-&$-1.35^{+0.28}_{-0.29}$&$\sim1\times10^{51}$&$\sim5\times10^{52}$&$\sim2\times10^{-5}$&$\sim1\times10^{4}$&$1.9\times10^{-2}$&75\\
130603B&(6)&0.359&-&$-1.63^{+0.35}_{-0.43}$&$5.0\times10^{49}$&$\sim1\times10^{51}$&$5\times10^{-3}$-$30$&$\sim4\times10^{3}$&$4.8\times10^{-2}$&10-31\\
140903A&(7)&0.351&-&-&$7.1\times10^{48}$&$4.3\times10^{52}$&$3.2\times10^{-2}$&$7.9\times10^{3}$&$1.7\times10^{-4}$&30\\
160821B&(8)&0.16  &-&$-1.76^{+0.13}_{-0.18}$&$8.2\times10^{49}$&$2.5\times10^{50}$&$1.6\times10^{-3}$&$\sim9\times10^{3}$&$2.5\times10^{-1}$&21\\
&(9)&&&&&$1.6\times10^{52}$&$10^{-4}$&$2.6\times10^{5}$&$5.1\times10^{-3}$&14\\
\hline
\multicolumn{11}{l}{{\bf Refs.} (1) \cite{Berger+2005}, (2) \cite{Soderberg+2006c}, (3) \cite{Fong+2015}, (4) \cite{Mangano+2007}, (5) \cite{Xu+2009},}\\
\multicolumn{11}{l}{(6) \cite{Fong+2014}, (7) \cite{Troja+2016}, (8) \cite{Troja+2019b}, (9) \cite{Lamb+2019}.}\\
\end{tabular}
\end{center}
\end{table*}

We calculate $\eta_{\rm EE}$ and $\Gamma_{\rm EE}$ by using Eqs. \eqref{eq eta} and \eqref{eq  gamma ee} with the observable quantities shown in Table \ref{table list2}.
We take the values of $E_{\rm EE,iso}$ from Table 2 in \cite{Kagawa+2019} and multiply a factor $6$ to convert the radiation energy measured in a band of 2-10 keV to that in 2-150 keV.
This conversion factor is given by assuming that EEs have a Band-like broken power-law spectrum \citep{Band+1993} with a peak energy $15\,\rm keV$ and low and high-energy spectral index of $-1$ and $-2$.\footnote{Note that in \S \ref{compactness} we used single power-law spectra with an exponential cutoff for EE's spectra to obtain conservative limits on the minimal Lorentz factor. The prescriptions are not inconsistent with each other.}
This prescription may be justified in particular for GRBs 050724, 060614, and 070714B because their photon indexes in BAT- ($15-150\,\rm keV$, during the EEs) and XRT-bands ($0.2-10\,\rm keV$, at the end of the EEs) are measured as $\sim-2$ and $\sim-1$, respectively.
In Table \ref{table list2}, we show the indexes taken from \cite{Kagawa+2019}.
Although we use the same correction factor for other GRBs, we should keep in mind that there may be systematic errors in the value of $E_{\rm EE,iso}$ due to our ignorance of EE's spectrum.
For GRBs 060313 and 111121A, which are not analyzed by \cite{Kagawa+2019}, we estimate $E_{\rm EE,iso}$ from the light curves shown in \cite{Kisaka+2017} taking the energy correction (multiplying a factor 3, see \citealt{Kagawa+2019}) into account, and assume their redshift as $z=0.5$.
For GRBs 051221A, 060614, 140903A, and 160821B, the durations of PEs are given explicitly in the references.
For the other GRBs, we estimate the duration from the light curves in Fig. \ref{fig lc} by eye inspection.

The data quality of the SGRBs used here, while being better than that for other SGRBs, is still too limited to obtain all of the SGRB parameters from afterglow analyses.
For example, the ISM density of GRB 130603B is poorly constrained within 4 orders-of-magnitude (Table \ref{table list2}).
For GRB 160821B, \cite{Lamb+2019} and \cite{Troja+2019b} constructed a multi-wavelength model independently, but their interpretations of the light curve such as a jet break do not match with each other.
We show both of their results in Table \ref{table list2} and use their estimated parameters equally.
We also comment on the analysis by \cite{Fong+2015}, where they fit X-ray light curves by a single power-law function only after $1000\,\rm s$ to avoid including EEs.
Although their analysis gives reasonable fits as shown in Fig. \ref{fig lc} (the slope of the blue lines), the fitting parameters may change if the PEs are taken into account in the analysis.
Thus, it should be noted that in addition to $E_{\rm EE,iso}$ the afterglow parameters obtained by these limited data may result in a systematic error for the values of $\eta_{\rm EE}$ and $\Gamma_{\rm EE}$.

\begin{figure}
\begin{center}
\includegraphics[width=85mm, angle=0]{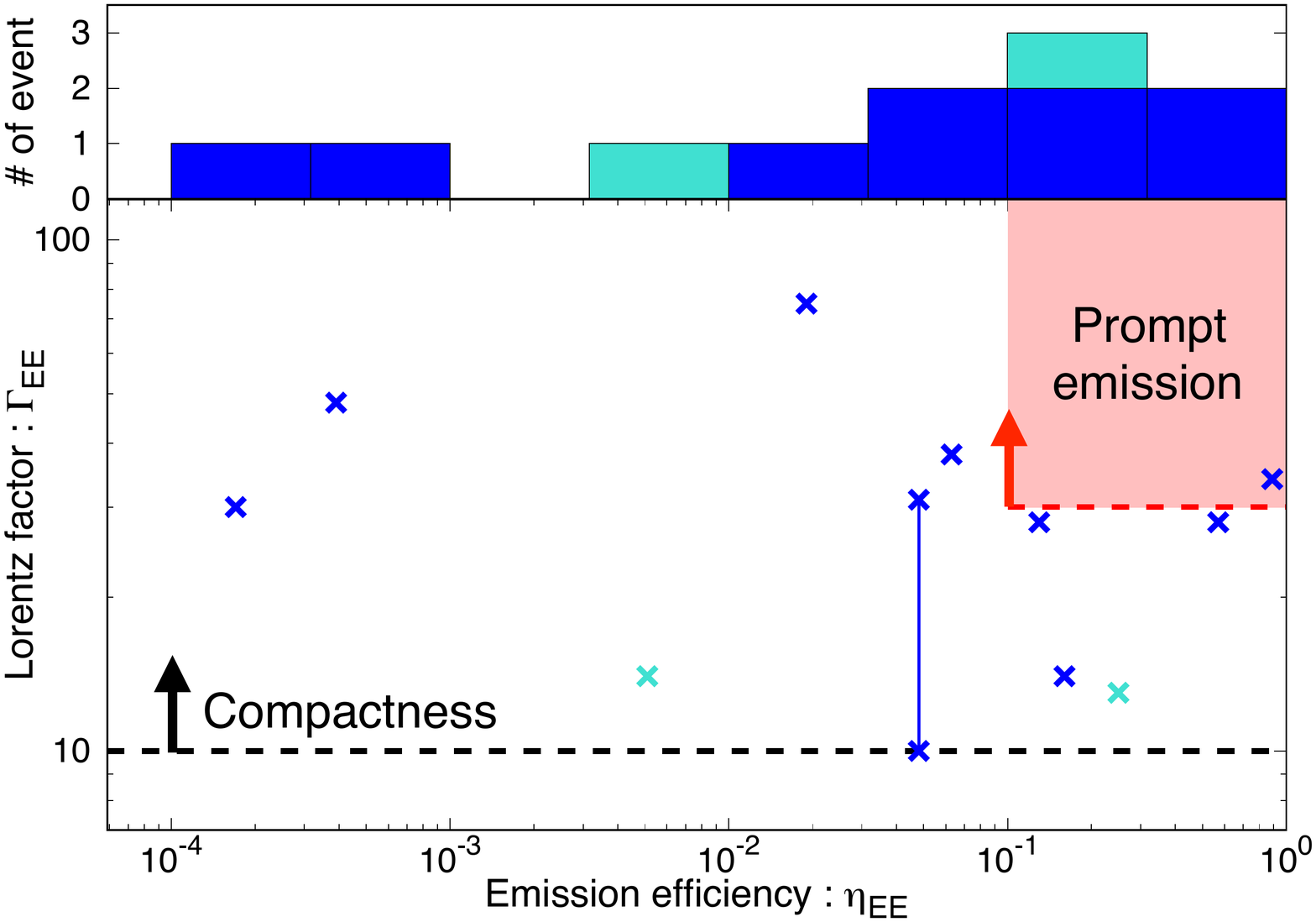}
\caption{({\bf Top}) The distribution of the emission efficiency of EEs, $\eta_{\rm EE}$ given by Eq. \eqref{eq eta}.
The mean, standard deviation, and median of $\log\eta_{\rm EE}$ are $-1.5$ ($\eta_{\rm EE}\sim3\,\%$), $1.2$, and $-1.2$ ($\eta_{\rm EE}\sim6\,\%$), respectively.
The light-blue bars show the efficiency for GRB 160821B given by independent analyses of \citealt{Lamb+2019} and \citealt{Troja+2019b} (see also Table \ref{table list2}). 
({\bf Bottom}) The $\eta_{\rm EE}$ - $\Gamma_{\rm EE}$ plot.
The Lorentz factors $\Gamma_{\rm EE}$ are consistent with the minimal Lorentz factor $\Gamma_{\rm min}\simeq10$ given by compactness (above the black dashed line).
The pink shaded region shows the parameter space of prompt emissions.
The blue points connected with a blue line represent GRB 130603B whose ISM density is not constrained tightly.}
\label{fig eta}
\end{center}
\end{figure}

The top panel of Fig. \ref{fig eta} depicts the distribution of the radiation efficiency of EE, $\eta_{\rm EE}$ calculated by Eq. \eqref{eq eta}.
Note that we use $E_{\rm EE,iso}$ evaluated for 2-150 keV, which thus gives a lower limit on the true radiation efficiency.
The mean and median of $\log\eta_{\rm EE}$ are $-1.5$ ($\eta_{\rm EE}=10^{-1.5}\sim3\,\%$) and $-1.2$ ($\eta_{\rm EE}=10^{-1.2}\sim6\,\%$), respectively.
These values are smaller than the emission efficiency of prompt emissions $\gtrsim10-50\,\%$ \citep{Fong+2015}, which may suggest a different emission mechanism of EEs from prompt emissions.
Although the number of events is small, the efficiency has a broad distribution with a standard deviation in $\log\eta_{\rm EE}$ of $1.2$.

We also plot the efficiencies with the Lorentz factors given by Eq. \eqref{eq gamma ee} in the bottom panel.
There seems to be no correlation between $\eta_{\rm EE}$ and $\Gamma_{\rm EE}$, although we may need more events to obtain a conclusive statement.
The Lorentz factors of all events satisfy the compactness limit $\Gamma_{\rm EE}\gtrsim10$ obtained in \S \ref{compactness} independently.
This implies that the slowest part of the EE-outflow contributes to produce the EE.
The pink shaded region represents the typical parameter space suggested for prompt emissions.
It should be noted that the minimal Lorentz factor of the SGRB prompt emission is constrained to be $\Gamma_0\gtrsim30$, which is smaller than that of long GRBs \citep{Nakar2007,Matsumoto&Piran2019}.
Some EEs have smaller Lorentz factors and radiation efficiency than those of the prompt emissions.
Combined with the fact that some SGRBs have more energetic EEs than prompt emissions \citep[$E_{\rm EE,iso}\gtrsim E_{\gamma,\rm iso}$,][]{Villasenor+2005,Perley+2009}, the results imply that EE-outflows may have much more kinetic energy than the prompt jets, which is consistent with our refreshed shock scenario for PEs.

Finally we comment that the Lorentz factors of EE-outflow $\Gamma_{\rm EE}$ and durations of PEs $T_{\rm PE}$ in Table \ref{table list2} are consistent with the time evolution of prompt jets expected in our scenario.
For GRBs 060614 and 160821B, the initial Lorentz factor $\Gamma_0$ can be estimated by identifying the onset of afterglow (the deceleration time $t_{\rm dec}$ of the prompt jet, e.g., \citealt{Nakar2007}).
We can calculate the Lorentz factor of the prompt-jet shell at the end of PE by assuming the adiabatic evolution, which should be smaller than $\Gamma_{\rm EE}$ in our scenario (of course, the deceleration time should be shorter than $T_{\rm PE}$).
For GRB 060614, \cite{Mangano+2007} regard a bump in the EE at $t_{\rm dec}\simeq40-50\,\rm s$ as an onset of afterglow and evaluate the Lorentz factor to be $\Gamma_0\simeq100$.
With this value, the Lorentz factor becomes $\simeq\Gamma_0(T_{\rm PE}/t_{\rm dec})^{-3/8}\simeq8$ at the end of the PE, which is smaller than $\Gamma_{\rm EE}$.
For GRB 160821B, \cite{Lamb+2019} also identify $t_{\rm dec}\simeq0.06\,\rm day$ and estimate $\Gamma_0\simeq55-60$. 
Noting that they also identify a jet break of the prompt jet at $t_{\rm j}\simeq0.35\,\rm day$, we estimate the Lorentz factor at the end of PE as $\simeq\Gamma_0(t_{\rm j}/t_{\rm dec})^{-3/8}(T_{\rm PE}/t_{\rm j})^{-1/2}\simeq10\lesssim\Gamma_{\rm EE}$. 
Although these estimations highly depend on the interpretation of the light curves, if correct, these inferred values are consistent with those expected in our picture.

\section{Summary and Discussion}\label{summary}
We showed that EEs should be produced by relativistic outflows and proposed that the outflows naturally power PEs by refreshed forward shocks formed when the slower ejecta responsible for  the EEs collide with the prior (initially faster but subsequently slowed-down) ejecta responsible for the prompt emission. 
By considering compactness arguments, which require that a \gr emitting region should be optically thin to scatterings between high-energy photons and electrons (and positrons), we calculated the minimal Lorentz factors of 39 EEs detected by BATSE, BAT, and GBM.
We found that EE-outflows --- outflows producing EEs --- should be relativistic with a Lorentz factor larger than $\Gamma\gtrsim10$. 
The EEs provide direct evidence that the central engine of SGRBs should maintain their activity for a long timescale $T_{\rm EE}\sim10^{2-3}\,\rm s$, and a magnetar or a BH can naturally continue to launch relativistic outflows over such timescales.
The EE-outflows can be less relativistic than the prompt jets, whose Lorentz factor are constrained to be $\Gamma_0\gtrsim30$ \citep{Nakar2007,Matsumoto&Piran2019}, because the EEs are less luminous than the prompt emissions. 

We also proposed a scenario for SGRBs accompanied by both EE and PE, where the PE is powered by a forward shock of decelerating prompt jet with an energy injection from an interaction with the EE-outflow.
Note that a similar scenario is discussed by \cite{Xu+2009,Lamb+2019} in the contexts of multi-wavelength modeling of specific SGRBs, but here we propose that this picture holds for all SGRBs with EE and PE, and in addition, for 10 well-observed SGRBs, we derived generic constraints on the emission efficiency $\eta_{\rm EE}$ and the bulk Lorentz factor $\Gamma$.
The interpretation that PEs are powered by refreshed forward shocks has several advantages over internal dissipation models. 
Long-lasting central-engine activities are not necessarily required, and a smooth transition from a PE to a normal decay can naturally be explained without a spectral evolution.

The efficiency shows a broad distribution with a standard deviation of $\log \eta_{\rm EE}=1.2$ and the mean and median values are smaller than the efficiency of prompt emissions $\gtrsim10-50\,\%$ \citep{Fong+2015}, which may imply a different emission mechanism from that of prompt emissions.
The derived Lorentz factor of EEs are consistent with compactness considerations $\Gamma_{\rm EE}\gtrsim10$.
There seems no correlation between $\eta_{\rm EE}$ and $\Gamma_{\rm EE}$, although the number of sample is limited.
More events will be needed to make decisive statements on the distributions.

Our results constrain or have implications on theoretical models of EEs.
First of all, as a model-independent limit, EEs should be powered by relativistic outflows with $\Gamma\gtrsim10$.
This condition may be satisfied by relativistic magnetar winds and relativistic BH jets as proposed by many authors.
On the other hand, \cite{Gompertz+2014,Gibson+2017} considered that an EE is powered by an outflow launched from a magnetar and a surrounding accretion disk system by the propeller effect, but it is unlikely that the propeller effect produces a relativistic outflow with $\Gamma\gtrsim10$ and this model may be disfavored.
\cite{Rezzolla&Kumar2015} proposed that EE is produced by an interaction between non-relativistic outflows, which is ruled out by the compactness arguments.

The mean and median values of $\eta_{\rm EE}\sim3-6\,\%$ are similar to those predicted by the internal shock model, as originally estimated for the prompt emission \citep{Kobayashi+1997,Daigne&Mochkovitch1998,Panaitescu+1999,Kumar1999}.
However, in order to reproduce the broad distribution of the radiation efficiency, the Lorentz factor profile of EE-outflows should be very different from event to event.
We note that for a magnetar model, \cite{Gompertz+2014} evaluated the necessary emission efficiency to explain EEs by the propeller effect as $\eta_{\rm EE}\gtrsim10\,\%$. Although this model is already made unlikely due to the compactness arguments, the required efficiency is also inconsistent with our result.

Our refreshed shock scenario requires that the kinetic energy of prompt jets should be smaller than that of EE-outflows, which implies a much larger emission efficiency of the prompt emission than that in previous estimates \citep[$\gtrsim10-50\,\%$, e.g., ][]{Fong+2015}.
If an EE-outflow has an energy distribution of $E(>\Gamma)\propto \Gamma^{1-s}$ after producing an EE,\footnote{We follow the notation in \cite{Sari&Meszaros2000}, where the mass distribution is given by $M(>\Gamma)\propto \Gamma^{-s}$.} the increase of the kinetic energy during the PE is given by $E(t)=E_{\rm k,iso}(t/T_{\rm PE})^{\frac{3(s-1)}{s+7}}$ ($t<T_{\rm PE}$), where we normalize $E(t)$ at the end of the energy injection.
The index $s$ is related with the temporal index $\alpha$ of the PE flux, $F_\nu\propto t^{\alpha}$, depending on the cooling regimes \cite[see, e.g.,][]{Sari&Meszaros2000}: 
\begin{align}
s=\begin{cases}
\frac{6p-3+7\alpha}{3-\alpha}&\text{; slow cooling,}\\
\frac{6p-2+7\alpha}{2-\alpha}&\text{; fast cooling,}
\end{cases}
\end{align}
respectively, where $p$ is the power-law index of the energy distribution of electrons.
For the typical value of $p=2.4$ \citep{Fong+2015} and $-0.5\leq\alpha\leq0$, the index becomes $s\simeq2-6$ and the kinetic energy evolution is given by $E(t)\propto t^{0.3-1.2}$.
Thus, for a PE continuing from $t\sim10^3\,\rm s$ to $T_{\rm PE}\sim10^{4}\,\rm s$, the original kinetic energy of the prompt jet is evaluated to be $10^{-(0.3-1.2)}\simeq0.06-0.5$ times smaller than that obtained by the normal decay.
The significant reduction of the original kinetic energy requires a large efficiency of prompt emissions, as pointed out for long GRBs with early-time shallow decays by e.g., \cite{Nousek+2006,Ioka+2006}.

One possible way in which this difference in the EE and prompt emission efficiencies may occur is if they involve different radiation mechanisms.
For example, the prompt emission may originate from the photospheric emission with high radiation efficiency \citep[e.g.,][]{Rees&Meszaros2005, Beloborodov2010}, whereas the EE can be attributed to synchrotron emission from non-thermal electrons produced in the relativistic outflow. We note that the energy injection from a long-lasting central engine or a delayed outflow refreshes not only the external forward shock but also the reverse shock. However, the latter component may be overwhelmed by the forward shock component, especially if the flows are significantly magnetized \citep{Zhang&Kobayashi2005}.

Late-time engine activities can be an energy source of kilonova/macronova emissions \citep[e.g.,][]{Kisaka+2015,Kisaka+2016,Matsumoto+2018b}, whose main energy source is canonically considered to be radioactive decay heating of \textit{r}-process elements \citep{Li&Paczynski1998,Kulkarni2005,Metzger+2010}.
If a central engine continues to launch a relativistic outflow and powers EEs, the outflow may form shocks and heat up ejecta, resulting in kilonova/macronova emission.
The previous works assumed that PEs are also powered by the internal dissipation of such outflows (not from external shocks), which is more favored than EEs because the energy injection at late times does not suffer from adiabatic losses.
However, as we proposed, PEs cannot have internal origins if the PEs are powered by refreshed shocks and the engines have already shut down at $\sim T_{\rm PE}$.
Further checks whether PEs are powered by the dissipation of jet-like outflows or not may be done by detecting early-time X-ray counterparts of NS mergers without prompt $\gamma$-ray emission \citep{Matsumoto&Kimura2018} or searching for nebular emission formed by the central engine \citep{Murase+2018}.
Finally, the prolonged engine activities in the PE phase can be also probed by long-lasting high-energy $\gamma$-ray and perhaps neutrino counterparts \citep{Kimura+2017,Murase+2018,Kimura+2019}. 

\section*{acknowledgments}
We thank Yasuaki Kagawa for kindly giving us the data of spectral indexes of extended emissions in his paper.
T.M. thanks Tsvi Piran and Kunihito Ioka for helpful comments.
This work is supported in part by JSPS Postdoctral Fellowship, Kakenhi Nos. 19J00214 (T.M.), 18H01245, 18H01246, and 19K14712 (S.S.K), Fermi GI program 111180 (K.M. and S.S.K.), NSF grant No. PAST-1908689 and the Alfred P. Sloan Foundation (K.M.), and the Eberly Foundation (P.M.). 

\bibliographystyle{mnras}
\bibliography{reference_matsumoto}

\appendix
\section{Sample}
We show the sample used to derived the histogram in Fig. \ref{fig gamma} in Table \ref{table list}.

\begin{table*}
\begin{center}
\caption{Sample of EEs used to derive the minimal Lorentz factors of EE-outflows (taken from \citealt{Bostanci+2013,Kaneko+2015}).
The first, second, and third groups are events detected by BATSE, BAT, and GBM, respectively. For redshift-unknown events, we assume $z=0.5$.
}
\label{table list}
\begin{tabular}{lrrccccrr}
\hline
Event&Fluence $\times10^{-7}$&Duration&Redshift&Spectrum&Peak energy&Spectral index&\multicolumn{2}{c}{Limit : $\Gamma_{\rm min}$}\\
&$S$ [$\rm erg\,cm^{-2}$]&$T$ [s]&$z$&&$\varepsilon_{\rm pk}$ [keV]&$\alpha_p$&Limit B&Limit C\\
\hline\hline
GRB 910725&$11.25\pm1.37$&86  &-&PL&-&$-2.57\pm0.26$&\bf 8.9&6.8\\
GRB 911016&$8.45\pm1.91$&112  &-&PL&-&$-1.42\pm0.23$&\bf 15.9&6.2\\
GRB 911119&$10.07\pm1.30$&81   &-&PL&-&$-2.06\pm0.19$&\bf 11.7&6.7\\
GRB 950531&$7.01\pm2.38$&55    &-&PL&-&$-2.77\pm0.65$&\bf 7.9&6.7\\
GRB 951211&$5.07\pm0.73$&38    &-&PL&-&$-2.65\pm0.28$&\bf 8.5&6.8\\
GRB 960906&$7.58\pm1.39$&87    &-&PL&-&$-1.64\pm0.27$&\bf 14.7&6.3\\
GRB 961017&$1.62\pm0.63$&45    &-&PL&-&$-1.19\pm0.31$&\bf 16.3&5.5\\
GRB 970918&$0.92\pm0.37$&15    &-&PL&-&$-1.99\pm0.68$&\bf 11.5&5.9\\
GRB 980112&$12.84\pm2.70$&82   &-&PL&-&$-1.99\pm0.34$&\bf 12.7&7.0\\
GRB 980904&$8.96\pm2.76$&111   &-&PL&-&$-2.59\pm0.67$&\bf 8.5&6.3\\
\hline
GRB 050724&$7.3\pm0.9$   &107 &0.26  &PL&-&$-2.04\pm0.18$&2.0&\bf 4.5\\
GRB 051016B&$1.3\pm0.3$ &33  & 0.94  &PL&-&$-2.87\pm0.57$&2.4&\bf 7.4\\
GRB 060614&$115.3\pm1.2$&169& 0.13  &PL&-&$-2.10\pm0.02$&1.9&\bf 5.0\\
GRB 061006&$8.6\pm0.8$    &113& 0.44  &PL&-&$-2.23\pm0.16$&2.2&\bf 5.9\\
GRB 061210&$9.2\pm1.2$    &77  & 0.41  &PL&-&$-1.73\pm0.20$&2.5&\bf 6.1\\
GRB 070506&$0.6\pm0.2$    &15  & 2.31  &PL&-&$-2.88\pm0.71$&3.8&\bf 11.2\\
GRB 070714B&$1.8\pm0.4$  &39  & 0.92  &PL&-&$-2.33\pm0.34$&2.8&\bf 7.5\\
GRB 080503&$19.4\pm0.9$  &147 &-        &PL&-&$-1.93\pm0.08$&2.6&\bf 6.8\\
GRB 090531B&$5.0\pm0.1$  &54  &-         &PL&-&$-1.79\pm0.17$&2.6&\bf 6.4\\
GRB 090927&$1.9\pm0.6$    &28  &1.37    &PL&-&$-1.98\pm0.45$&3.9&\bf 9.9\\
GRB 100212A&$6.1\pm0.8$ &135 &-         &PL&-&$-2.47\pm0.21$&2.1&\bf 5.7\\
GRB 100522A&$6.1\pm0.3$ &15   &-         &PL&-&$-2.56\pm0.10$&2.3&\bf 8.2\\
GRB 110207A&$17.0\pm1.1$&137 &-        &PL&-&$-1.46\pm0.09$&2.9&\bf 6.7\\
GRB 110402A&$29.7\pm2.7$&82   &-        &PL&-&$-1.93\pm0.14$&2.8&\bf 8.0\\
GRB 111121A&$11.8\pm0.7$ &61  &-        &PL&-&$-2.02\pm0.10$&2.6&\bf 7.2\\
GRB 121014A&$9.7\pm0.8$  &81  &-        &PL&-&$-2.00\pm0.14$&2.5&\bf 6.7\\
\hline
GRB 080807&$18.5\pm0.9$&27   &-&PL&-&$-1.33\pm0.04$&5.4&\bf 8.9\\
GRB 090131&$49.6\pm0.7$&23   &-&CPL&$54.5\pm14.0$&$-1.49\pm0.03$&5.6&\bf 10.8\\
GRB 090820&$7.2\pm0.5$&7       &-&PL&-&$-2.22\pm0.06$&4.5&\bf 9.6\\
GRB 090831&$112.2\pm3.4$&86 &-&CPL&$532\pm422$&$-1.65\pm0.06$&\bf 16.5&9.9\\
GRB 091120&$192.3\pm2.5$ &52&-&CPL&$114.4\pm3.9$&$-1.10\pm0.03$&3.2&\bf 11.8\\
GRB 100517&$14.0\pm0.8$&11   &-&CPL&$26.8\pm3.2$&$-1.39\pm0.19$&1.3&\bf 9.9\\
GRB 100522A&$9.9\pm0.8$&13     &-&PL&-&$-2.21\pm0.07$&4.4&\bf 9.1\\
GRB 110207A&$22.3\pm1.5$&38   &-&PL&-&$-1.23\pm0.06$&5.5&\bf 8.7\\
GRB 110402A&$58.5\pm3.1$&39   &-&PL&-&$-1.43\pm0.05$&5.5&\bf 10.2\\
GRB 110824&$69.4\pm2.2$&93   &-&PL&-&$-1.59\pm0.03$&5.1&\bf 9.1\\
GRB 120402&$4.7\pm0.8$&19    &-&PL&-&$-2.12\pm0.13$&4.2&\bf 7.5\\
GRB 120605&$6.4\pm1.6$&8      &-&PL&-&$-1.98\pm0.06$&4.7&\bf 9.2\\
GRB 121029&$52.6\pm0.7$&6    &-&CPL&$178.6\pm6.1$&$-0.34\pm0.06$&3.2&\bf 13.7\\
\hline
\end{tabular}
\end{center}
\end{table*}

\bsp
\label{lastpage}
\end{document}